\pdfoutput=1

\documentclass[11pt,twoside,a4paper,cmspaper,final,collab]{cms-tdr}
\def\svnVersion{82004be-D}\def\svnDate{2019/12/11}\def\cmsCernNoTag{CERN-EP-2017-180}\def\cmsCernDate{\today}\def\cmsMessage{"Published in Physical Review C as \href{http://dx.doi.org/10.1103/PhysRevC.100.064908}{\doi{10.1103/PhysRevC.100.064908}.}"}

\begin{document}\cmsNoteHeader{HIN-16-017}

\hyphenation{had-ron-i-za-tion}
\hyphenation{cal-or-i-me-ter}
\hyphenation{de-vices}
\RCS$Revision$
\RCS$HeadURL$
\RCS$Id$
\newlength\cmsFigWidth
\ifthenelse{\boolean{cms@external}}{\setlength\cmsFigWidth{0.98\columnwidth}}{\setlength\cmsFigWidth{0.65\textwidth}}
\ifthenelse{\boolean{cms@external}}{\providecommand{\cmsLeft}{top\xspace}}{\providecommand{\cmsLeft}{left\xspace}}
\ifthenelse{\boolean{cms@external}}{\providecommand{\cmsRight}{bottom\xspace}}{\providecommand{\cmsRight}{right\xspace}}
\ifthenelse{\boolean{cms@external}}{\providecommand{\cmsTable}[1]{#1}}{\providecommand{\cmsTable}[1]{\resizebox{\textwidth}{!}{#1}}}
\providecommand{\sqrtsNN}{\ensuremath{\sqrt{\smash[b]{s_{_{\mathrm{NN}}}}}}\xspace}
\newcommand{\Dzerodecay}{\ensuremath{\PDz\to \PKp\PGpp}\xspace}
\newcommand {\PbPb}  {\ensuremath{\mathrm{PbPb}}\xspace}
\newcommand {\pPb}  {\ensuremath{\Pp\mathrm{Pb}}\xspace}
\newcommand {\AuAu}  {\ensuremath{\mathrm{AuAu}}\xspace}
\newcommand {\AonA}  {\ensuremath{\mathrm{AA}}\xspace}
\newcommand{\noff}    {\ensuremath{N_\text{trk}^\text{offline}}\xspace}
\ifthenelse{\boolean{cms@external}}{\providecommand{\NA}{\ensuremath{\cdots}}}{\providecommand{\NA}{\ensuremath{\text{---}}}}

\cmsNoteHeader{HIN-16-017}
\title{Probing the chiral magnetic wave in \texorpdfstring{\pPb and \PbPb collisions at $\sqrtsNN = 5.02\TeV$}{pPb and PbPb collisions at sqrt(s[NN]) = 5.02 TeV} using charge-dependent azimuthal anisotropies}

\date{\today}

\abstract{
Charge-dependent anisotropy Fourier coefficients ($v_n$)
of particle azimuthal distributions are measured in \pPb and
\PbPb collisions at $\sqrtsNN = 5.02\TeV$ with the CMS
detector at the LHC. The normalized difference in the second-order anisotropy coefficients ($v_2$) between positively and negatively charged particles is found to depend linearly on the observed event charge asymmetry with comparable slopes for both \pPb and \PbPb collisions over a wide range of charged particle multiplicity. In \PbPb, the third-order anisotropy coefficient, $v_3$, shows a similar linear dependence with the same slope as seen for $v_2$. The observed similarities between the $v_2$ slopes for \pPb and \PbPb, as well as the similar slopes for $v_2$ and $v_3$ in \PbPb, are compatible with expectations based on local charge conservation in the decay of clusters or resonances, and constitute a challenge to the hypothesis that, at LHC energies, the observed charge asymmetry dependence of $v_2$ in heavy ion collisions arises from a chiral magnetic wave.
}

\hypersetup{%
pdfauthor={CMS Collaboration},%
pdftitle={Probing the chiral magnetic wave using charge-dependent azimuthal anisotropies in pPb and PbPb collisions at sqrt(s[NN]) = 5.02 TeV},%
pdfsubject={CMS},%
pdfkeywords={CMS, physics, CMW, small systems}}

\maketitle

\section{Introduction}

Observing macroscopic phenomena arising from quantum anomalies is a subject of interest for a wide range of physics communities, from magnetized relativistic matter in three-dimensional
Dirac and Weyl materials~\cite{Lv:2015pya,Huang:2015eia,Li:2014bha} to hot plasma in the early
universe or formed in relativistic heavy ion collisions~\cite{Abelev:2009ad,Abelev:2012pa,Khachatryan:2016got}.
In quantum chromodynamics, gluon fields within a localized region of
space-time can form nontrivial topological
configurations~\cite{Lee:1973iz,Lee:1974ma,Morley1985,Kharzeev:1998kz}.
If approximate chiral symmetry is restored,
the interactions of chiral quarks with these gluon fields can produce a chirality imbalance,
violating the local $P$ and $CP$ symmetries~\cite{Morley1985,Kharzeev:1998kz}.
This anomalous chiral effect
can manifest itself as an electric current along or opposite
to a strong magnetic field~\cite{Kharzeev:2004ey,Kharzeev:2007jp,Kharzeev:2015znc}.
The electric charge separation produced by these currents is known as the chiral magnetic
effect (CME)~\cite{Kharzeev:2004ey}. The chiral separation effect
(CSE) is a similar process, where the separation of the chiral charges
along the magnetic field will be induced by a finite density of the net electric charges~\cite{Burnier:2011bf}.
The coupling of electric and chiral charge densities and currents leads to a long-wavelength collective excitation, known as the chiral magnetic wave (CMW)~\cite{Burnier:2011bf, Newman:2005hd, Kharzeev:2010gd, Gorbar:2011ya}.

In relativistic heavy ion (\AonA) collisions, a strong magnetic field and the restoration of the approximate chiral symmetry, both necessary conditions for creating a CMW, may be present. The magnetic field is produced by the spectator protons and is, on average, perpendicular to the reaction plane defined by the impact parameter and beam directions. The propagation of the CMW leads to an electric quadrupole moment,
where additional positive (negative) charges are accumulated away
from (close to) the reaction plane~\cite{Burnier:2011bf}.
Following a hydrodynamic evolution of the medium formed in \AonA\ collisions,
this electric quadruple moment is expected to result in a charge-dependent variation
of the second-order anisotropy coefficient ($v_2$) in the Fourier expansion of the final-state particle azimuthal distribution. More specifically,
the $v_{2}$ coefficient will exhibit a linear dependence on the observed event
charge asymmetry~\cite{Burnier:2011bf}, $A_\text{ch}\equiv (N_{+}-N_{-})/(N_{+}+N_{-})$,
where $N_{+}$ and $N_{-}$ denote the number of positively and negatively charged
hadrons in each event,
\begin{linenomath}
\begin{equation}
\label{f.ellipticflow}
v_{2,\pm} = v_{2,\pm}^\text{base} \mp rA_\text{ch}.
\end{equation}
\end{linenomath}
\noindent Here $v_{2,\pm}^\text{base}$ represents the value in the absence of a charge
quadrupole moment from the CMW for positively ($+$)
and negatively ($-$) charged particles,
and $r$ denotes the slope parameter. In the presence of a CMW, the difference of $v_2$ values between positively and negatively charged particles will be proportional to $A_\text{ch}$.
Similar charge-dependent effects from the CMW are not expected for
the third-order anisotropy coefficient ($v_3$)~\cite{Kharzeev:2015znc}.

Recent observations of the $A_\text{ch}$ dependence of $v_{2,\pm}$ in \AonA\ collisions
at RHIC at BNL and the CERN LHC are qualitatively consistent with expectations of the CMW
mechanism~\cite{Abelev:2012pa, Adamczyk:2015eqo, Adam:2015vje}. However, the interpretation
of the results remains inconclusive since
alternative mechanisms have been proposed to generate
charge-dependent $v_{2}$ coefficients without a CMW~\cite{Bzdak:2013yla,Hatta:2015hca}. For example, it has been shown that local charge conservation (LCC) in the decay of clusters or resonances can qualitatively describe
the charge-dependent $v_2$ data~\cite{Bzdak:2013yla}.
Decay particles from a lower transverse momentum
(\pt) resonance tend to have a larger rapidity separation, resulting in a
daughter more likely to fall outside the detector acceptance, leading to a nonzero
$A_\text{ch}$. Hence, this process generates a correlation between $A_\text{ch}$ and the average \pt
of charged particles, and therefore also between $A_\text{ch}$ and the $v_{2}$ coefficient, since $v_{2}$ depends on \pt.
The LCC mechanism also applies to all higher-order anisotropy Fourier coefficients ($v_n$).

This paper presents measurements of the $A_\text{ch}$ dependence of the
$\left< \pt \right>$ and of the $\pt$-averaged $v_{n}$ coefficients in \pPb and  \PbPb collisions at $\sqrtsNN = 5.02\TeV$, using
data collected with the CMS experiment at the LHC. It has been shown that $\Pp\Pp$ and \pPb
collisions with high charged-particle multiplicities can
generate large final-state azimuthal anisotropies, comparable to those
in \AonA\ collisions at similar event
multiplicities~\cite{Khachatryan:2010gv,Aad:2015gqa,Khachatryan:2015lva,Khachatryan:2016txc,CMS:2012qk,alice:2012qe,atlas:2012fa,Aaij:2015qcq,Khachatryan:2014jra,ABELEV:2013wsa,Chatrchyan:2013nka,Aad:2014lta,Khachatryan:2015waa,Dusling:2015gta}.
However, the CMW contribution to any $A_\text{ch}$-dependent $v_2$ signal is
expected to be negligible in \pPb collisions: the induced magnetic field is
smaller than in \PbPb\ collisions (albeit of the same order of magnitude) and,
more importantly, its correlation with the harmonic event planes is vanishingly small ~\cite{Khachatryan:2016got,Belmont:2016oqp}.
The recent observation of nearly identical charge-dependent azimuthal correlations in \pPb and \PbPb suggested significant contamination of background sources (e.g., LCC) to any CME induced signal~\cite{Khachatryan:2016got,Sirunyan:2017quh}.
Therefore, a comparison between \pPb and \PbPb systems and their $A_\text{ch}$ dependence of the $\left< \pt \right>$ and the $v_{3}$ coefficient can differentiate between the CMW and LCC mechanisms.
It is worth noting that a lack of experimental evidence for the CME~\cite{Khachatryan:2016got,Sirunyan:2017quh}
does not necessarily imply the absence of the CMW, as the CME requires an initial chirality
imbalance from topological QCD charges (which may be too weak to be observed),
whereas the CMW only requires an initial net electric charge density~\cite{Kharzeev:2010gd,Burnier:2011bf}.
Therefore, the CME and CMW deserve independent experimental investigations.

\section{The CMS detector}

The central feature of the CMS apparatus is a superconducting solenoid of 6\unit{m}
internal diameter, providing a magnetic field of 3.8\unit{T}. Within the solenoid volume,
there are silicon pixel and strip tracker detectors, a lead tungstate crystal electromagnetic calorimeter,
and a brass and scintillator hadron calorimeter. The silicon tracker
measures charged particles within the pseudorapidity range $\abs{\eta}< 2.5$. For charged
particles with $1 < \pt < 10\GeVc$ and $\abs{\eta} < 1.4$, the track resolutions are
typically 1.5\% in \pt and 25--90\,(45--150)\mum in the transverse (longitudinal)
impact parameter~\cite{TRK-11-001}. Iron and quartz-fiber Cherenkov hadron forward
(HF) calorimeters cover the range $2.9 < \abs{\eta} < 5.2$. A detailed description
of the CMS detector, together with a definition of the coordinate system used and
the relevant kinematic variables, can be found in Ref.~\cite{Chatrchyan:2008zzk}.

\section{Event and track selections}

The \pPb data at $\sqrtsNN = 5.02\TeV$, collected in 2013 using the CMS detector,
correspond to an integrated luminosity of 35\nbinv.
A subset of peripheral \PbPb data at $\sqrtsNN = 5.02\TeV$ collected
in 2015 (30--90\% centrality, where centrality is defined as the fraction of the total
inelastic cross section, with 0\% denoting the most central collisions~\cite{Chatrchyan:2012ta}), is also used. The sample is
reconstructed with the same algorithm as the \pPb data,
in order to compare directly the two systems at similar multiplicities.
The event reconstruction, event selection and the trigger,
including the dedicated triggers to collect a large sample of
high-multiplicity \pPb events, are identical to those used in previous CMS particle
correlation measurements~\cite{Chatrchyan:2013nka,Khachatryan:2010gv,Khachatryan:2016got}. In the offline
analysis of \pPb (\PbPb) collisions, hadronic events are selected by requiring the
presence of at least one (three) energy deposit(s) greater than 3\GeV in each of the
two HF calorimeters. Events are also required to contain a primary vertex within
15\unit{cm} of the nominal interaction point along the beam axis and 0.15\unit{cm}
in the transverse direction. In the \pPb data sample, there is a 3\% probability to
have at least one additional interaction in the same bunch crossing (pileup). After the procedure used to reject pileup events is applied, the remaining sample has a purity of 99.8\%
for single collision events~\cite{Chatrchyan:2013nka}.
The pileup in  \PbPb data is negligible.

Primary tracks, \ie, tracks that originate at the primary
vertex and satisfy the high-purity criteria of Ref.~\cite{TRK-11-001},
are used to define the event charged-particle multiplicity (\noff) and to perform correlation
measurements. In addition, the impact parameter significance of the tracks with
respect to the primary vertex in the beam and transverse direction is required to be less than 3. The relative uncertainty in \pt must be less than 10\%.
To ensure high tracking efficiency, only tracks with
$\abs{\eta}<2.4$ and $\pt > 0.3\GeVc$ are used for $A_\text{ch}$ and $v_{n}$ measurements in this analysis. The \pPb and \PbPb  data are compared in ranges of \noff, where primary tracks with $\abs{\eta}<2.4$ and $\pt >0.4$\GeVc are counted, in order to match the trigger selection criterion implemented at the HLT in \pPb collisions.

\begin{figure}[htb]
\centering
\includegraphics[width=\cmsFigWidth]{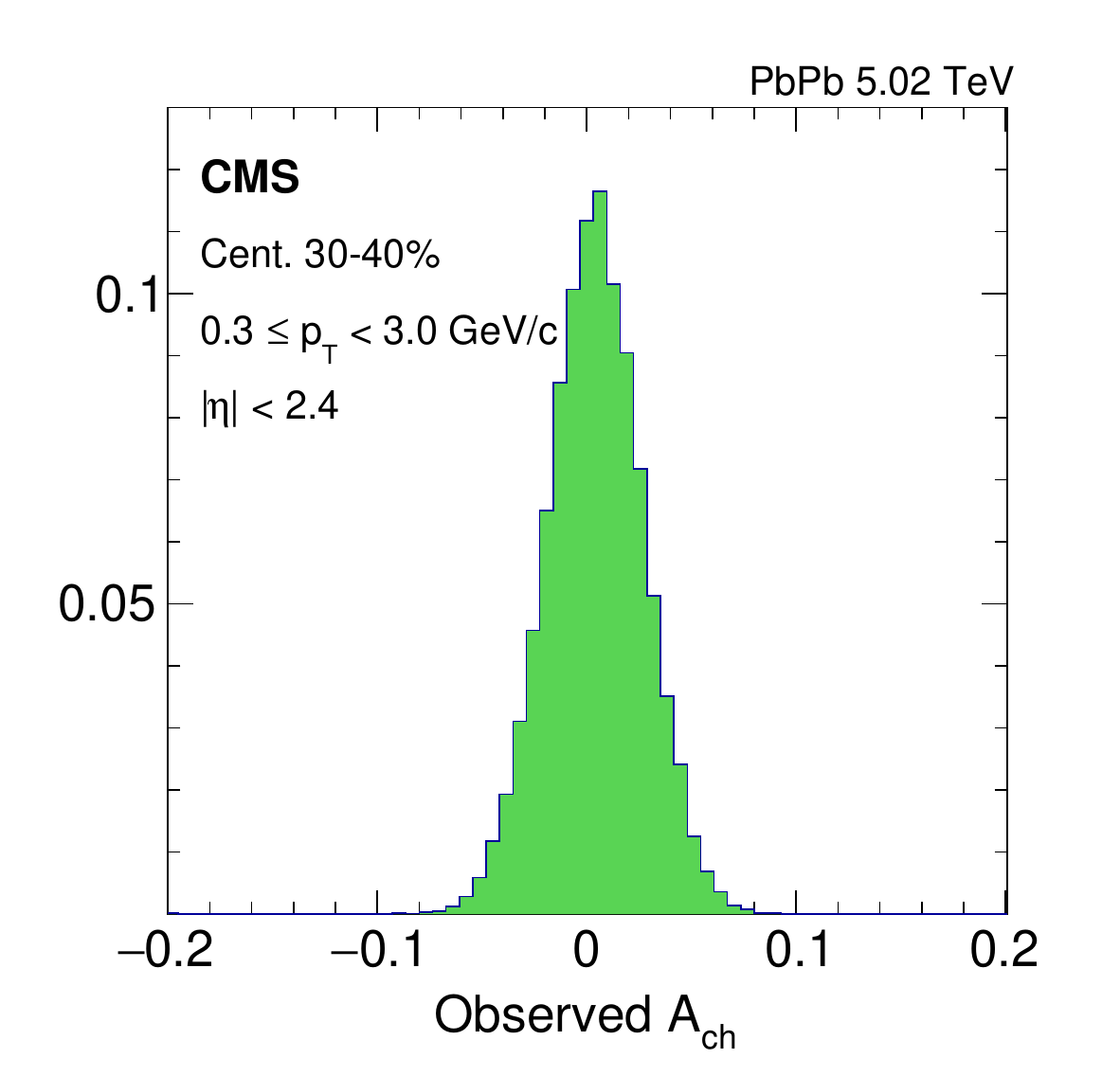}
\caption{
The event-by-event probability distribution observed in the charge asymmetry, $A_\text{ch}$, for \PbPb collisions at $\sqrtsNN = 5.02$\TeV\ within the 30--40\% centrality range. The particles are selected between 0.3 to 3.0\GeVc and having pseudorapidity $\abs{\eta} < 2.4$.}
\label{fig:Ach}
\end{figure}

\section{Analysis technique}

In each multiplicity or centrality class, events are further divided into several
ranges of the observed event charge asymmetry, $A^\text{obs}_\text{ch}$,
calculated based on the number of positively and negatively charged particles from primary tracks.
An example of the $A^\text{obs}_\text{ch}$ distribution for \PbPb data
in the 30--40\% centrality range is shown in Fig.~\ref{fig:Ach}.
Within each $A^\text{obs}_\text{ch}$ range, the $v_n$ coefficients are obtained
separately for tracks with positive ($v^{+}_n$) and negative ($v^{-}_n$) charge,
and with $\abs{\eta}<2.4$ and $0.3<\pt<3$\GeVc, using the two-particle cumulant
method~\cite{Bilandzic:2010jr} with a pseudorapidity gap of at least 1 unit between the two particles to
suppress the short-range correlations. Because of statistical limitations,
the pseudorapidity gap chosen in this analysis is smaller
than the value of 2 units typically used in other CMS correlation measurements, but results are found to be consistent between 1 and 2 units of pseudorapidity gap.
Residual effects of short-range correlations may still contribute
to the sum of the $v_{ n}$, $v^{-}_{n}+v^{+}_{n}$, but not the difference since the effect is largely canceled out.
However, this effect contributes to the \pPb and  \PbPb systems similarly~\cite{Chatrchyan:2013nka},
so it has little impact on the comparison of the two systems.

The main physics observable of interest in this analysis is the
slope parameter ($r^\text{norm}$) extracted by fitting a linear
function to the normalized $v_{n}$ differences,
$(v^{-}_{n}-v^{+}_{n})/(v^{-}_{n}+v^{+}_{n})$, as a function of
the true event charge asymmetry value, $A^\text{true}_\text{ch}$,
obtained by correcting $A^\text{obs}_\text{ch}$ for the detector acceptance
and tracking efficiency.
Based on Monte Carlo (MC) simulations, detector effects can be modeled as
a Gaussian response of the $A^\text{true}_\text{ch}$ distribution within $\abs{\eta}<2.4$, with a width
determined from the simulated $A^\text{obs}_\text{ch}$ distribution at a given $A^\text{true}_\text{ch}$ value.
Combining the $A^\text{obs}_\text{ch}$ distribution in data with the response function from MC simulations, the predicted correlation between $A^\text{obs}_\text{ch}$ and $A^\text{true}_\text{ch}$ in data is calculated. The slope of a linear fit to this correlation is used to obtain the average $A^\text{true}_\text{ch}$ value in each selected $A^\text{obs}_\text{ch}$ range in data. The slope, which ranges from 0.6 to 0.8, is fit separately for each multiplicity or centrality selection. This procedure is validated using different MC generators, which give similar correction factors.

\begin{figure}[htb]
\centering
\includegraphics[width=\linewidth]{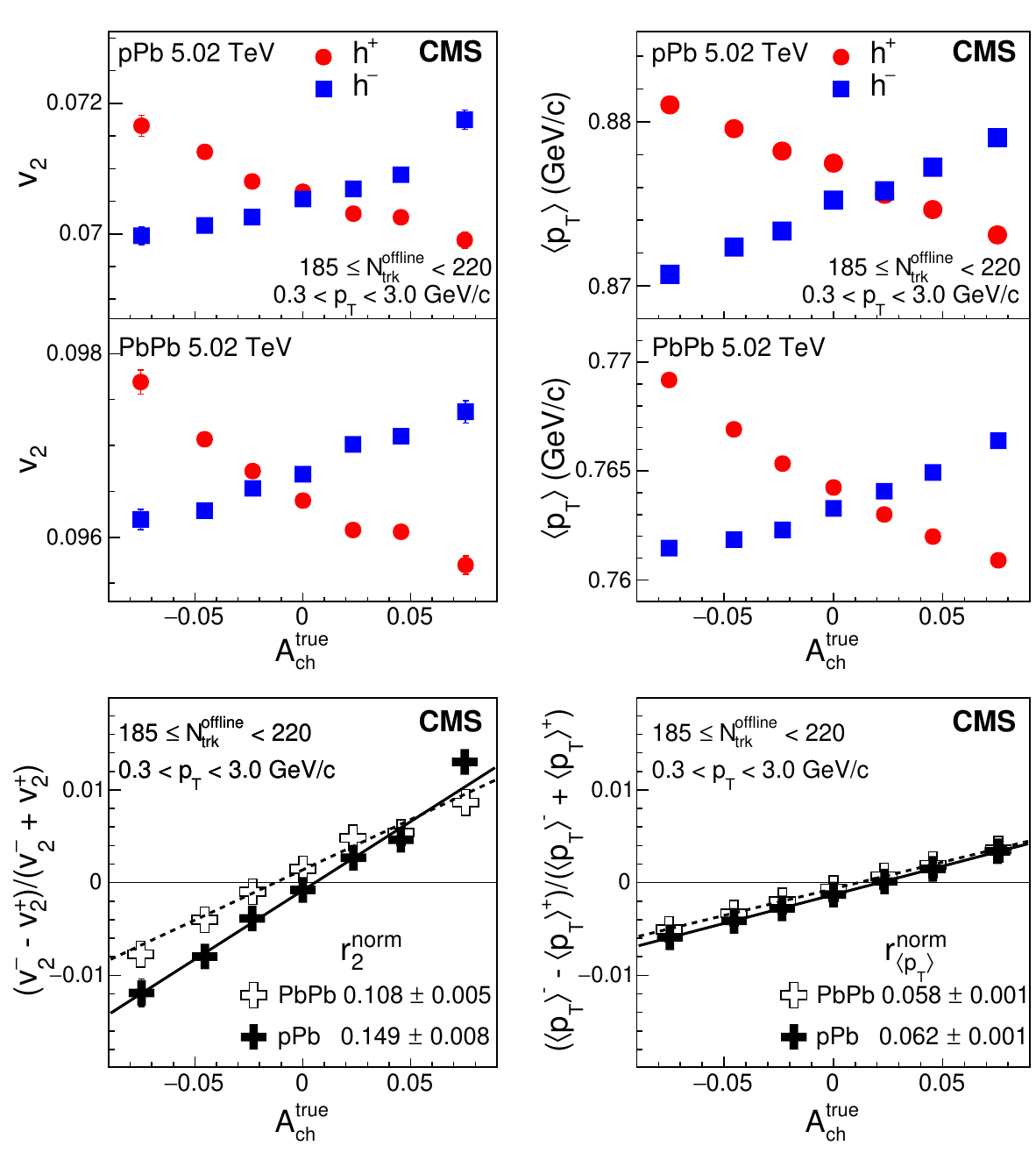}
\caption{
The elliptic anisotropy $v_2$ (top left) and event-averaged $\left< \pt \right>$ (top right)
for positively (h$^{+}$) and negatively (h$^{-}$) charged particles, and their normalized
differences (bottom row),
as functions of $A^\text{true}_\text{ch}$ for the multiplicity range $185 \leq \noff < 220$ of \pPb and \PbPb collisions
at $\sqrtsNN = 5.02\TeV$. Statistical uncertainties are smaller than the marker size, while
systematic uncertainties are not displayed.}
\label{fig:paperFigure_1}
\end{figure}

\begin{figure}[hbt]
\centering
\includegraphics[width=\cmsFigWidth]{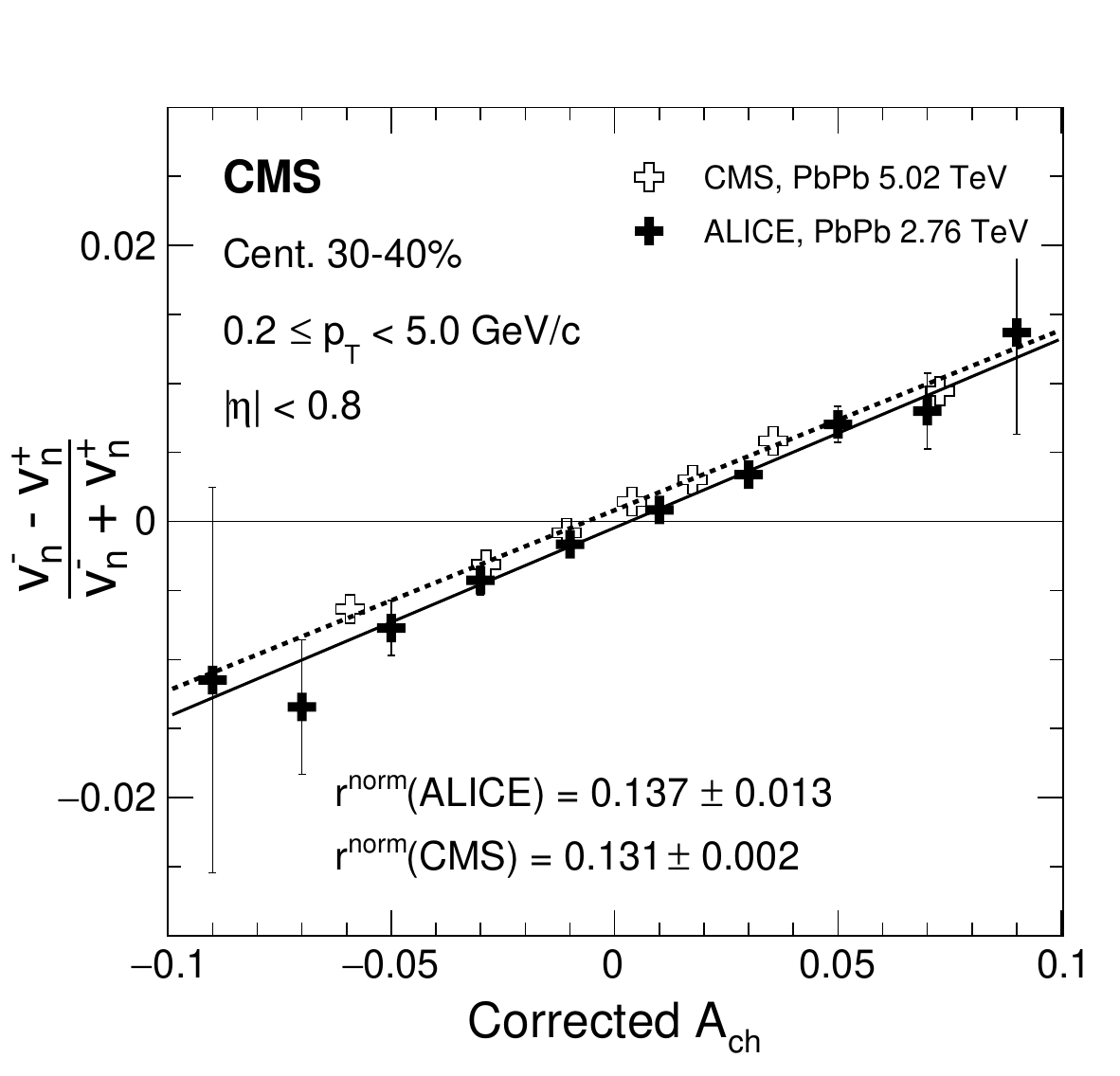}
\caption{
The normalized difference in elliptic flow $v_2$ between positive- and negative-charged particles, $(v^{-}_{2} - v^{+}_{2})/(v^{-}_{2} + v^{+}_{2})$, as a function of charge asymmetry, is presented. The results are selected in centrality range
30--40\% with particles within $\abs{\eta} < 0.8$ and $0.2 \leq \pt< 5.0 \GeV$, and are compared between the ALICE~\cite{Adam:2015vje} and the CMS experiment in \PbPb collisions at $\sqrtsNN = 2.76\TeV$ and 5.02\TeV, respectively. The bars represent statistical point-by-point uncertainties.}
\label{fig:figure_alicecomparison}
\end{figure}

The systematic uncertainty related to the $A_\text{ch}$ correction factors,
based on the difference between \textsc{epos lhc}~\cite{Pierog:2013ria} and \textsc{hydjet++}~\cite{Lokhtin:2008xi}
event generators, is estimated to be 1--7\% ranging from high- to low-multiplicity events.
To evaluate the systematic uncertainty related to the $v_{n}$ measurement,
the sensitivity of the results to different track selection criteria is studied.
Varying the longitudinal and transverse track impact parameter selection criteria from the default three standard deviations to two or five, and the relative \pt\ uncertainty selection criterion from the default 10\% to 5\%, yields a systematic uncertainty of less than 2\%.
The longitudinal primary vertex position ($z_\text{vtx}$) has been varied, using ranges
$\abs{z_\text{vtx}} < 3$\unit{cm} and $3 < \abs{z_\text{vtx}} < 15$\unit{cm}, where
the difference with respect to the default range $\abs{z_\text{vtx}} < 15$\unit{cm} is less than 2\%.
All of the systematic uncertainty sources are uncorrelated and were found to be similar for \pPb and  \PbPb collisions. Therefore, the total systematic uncertainty is taken as the quadratic
sum, and the same values are quoted for both \pPb and  \PbPb systems.

\section{Results}

Figure~\ref{fig:paperFigure_1} (left column) shows the $A^\text{true}_\text{ch}$ dependence
of $v_2$ coefficients, averaged over $0.3<\pt<3$\GeVc,
for positively and negatively charged particles in the multiplicity
range $185 \leq \noff < 220$ of \pPb and \PbPb collisions at $\sqrtsNN = 5.02$\TeV. The
normalized $v_2$ difference as a function of $A^\text{true}_\text{ch}$ is also shown.
A trend of $v^{+}_{2}$ ($v^{-}_{2}$) decreasing
(increasing) as $A^\text{true}_\text{ch}$ increases is observed for both \pPb and \PbPb collisions
with an approximately linear dependence.
A similar linear trend of elliptic anisotropy as a function of $A_\text{ch}$
has been observed in \AuAu~\cite{Adamczyk:2015eqo} and \PbPb~\cite{Adam:2015vje}
systems at lower collision energies, as shown in Fig.~\ref{fig:figure_alicecomparison} for 30--40\% centrality \PbPb events.
The linear slope parameter, $r^\text{norm}_{2}$, is extracted by a $\chi^{2}$
fit to a linear function, which gives values of $0.149 \pm 0.008$ for
\pPb and $0.108 \pm 0.005$ for \PbPb, in the multiplicity range $185 \leq \noff < 220$.
A significant nonzero value of the
linear slope parameter is observed in \pPb collisions, even greater than that
in \PbPb collisions.
Since the CMW effect is expected to be negligible in high-multiplicity \pPb events,
this observation might be caused, at LHC energies, by a mechanism unrelated to
the CMW. The differences in the linear slope parameters observed in the \pPb\ and \PbPb\
systems remain to be understood.

The $\left< \pt \right>$ for
positively and negatively charged particles are also measured as functions of
$A^\text{true}_\text{ch}$, in the multiplicity
range $185 \leq \noff < 220$ of \pPb and \PbPb collisions at $\sqrtsNN = 5.02$\TeV, and
shown in Fig.~\ref{fig:paperFigure_1} (right column). The normalized $\left< \pt \right>$
difference as a function of $A^\text{true}_\text{ch}$ is obtained for the two
systems with the slope parameters displayed in the figure. A similar linear
$A^\text{true}_\text{ch}$ dependence of the $\left< \pt \right>$ value to that of $v_2$ is
observed. This behavior is qualitatively consistent with the
expectation of the LCC effect from resonance decays.
Since $v_n$ has a strong dependence on particle \pt, a correlation between the
$\pt$-averaged $v_n$ and $A_\text{ch}$, as observed in Fig.~\ref{fig:paperFigure_1}
(left), can also be induced by the LCC mechanism.

\begin{figure}[htb]
\centering
\includegraphics[width=\cmsFigWidth]{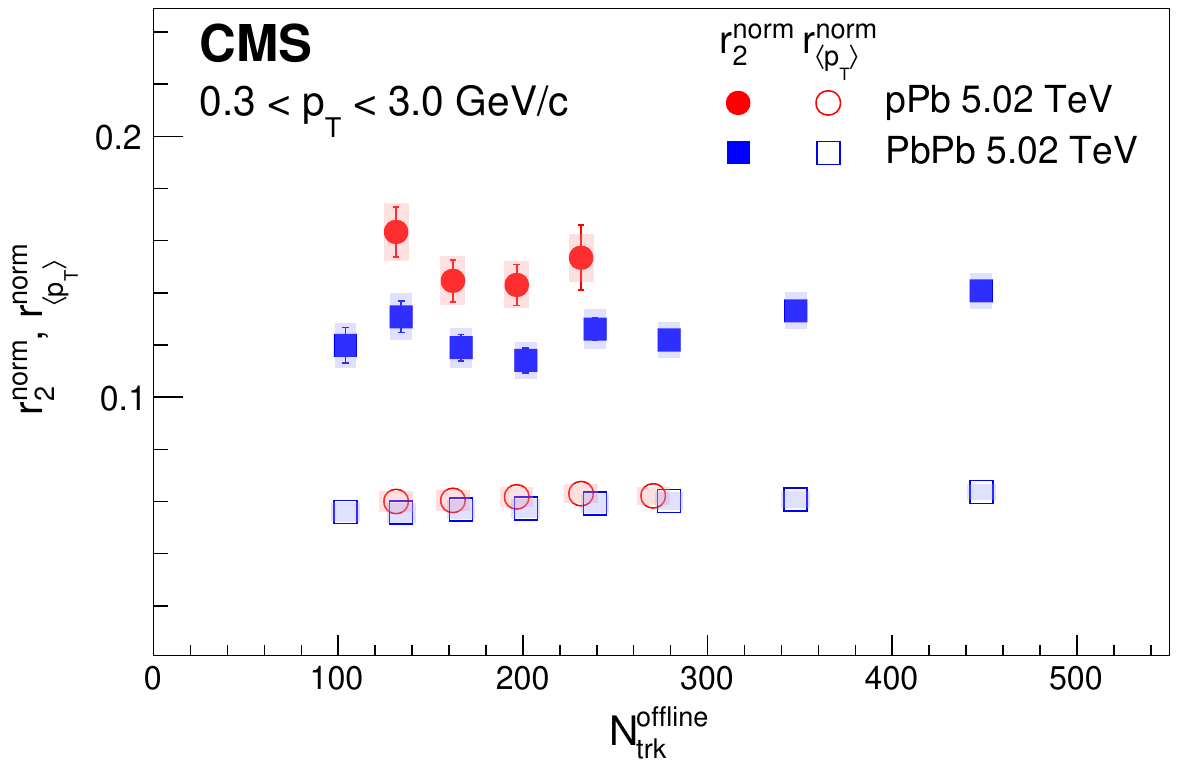}
\caption{
The linear slope parameters, $r^\text{norm}$, for $v_2$ (filled symbols) and
$\left< \pt \right>$ (open symbols) as functions of event multiplicity in \pPb and  \PbPb collisions at $\sqrtsNN = 5.02$\TeV.
Statistical and systematic uncertainties
are indicated by the error bars and shaded regions, respectively.}
\label{fig:figure2}
\end{figure}

The extracted normalized slope parameters for $v_2$ and $\left< \pt \right>$
as functions of event multiplicity in \pPb and
\PbPb collisions are shown in Fig.~\ref{fig:figure2}.
The $r^\text{norm}$ values for both $v_2$ and $\left< \pt \right>$ are found to have a weak
dependence on the event multiplicity for both \pPb and \PbPb collisions, with values for
$\left< \pt \right>$ approximately half of those for $v_2$.
In the overlapping multiplicity range, normalized slope
parameters are observed to be larger in \pPb than \PbPb collisions, which is
not expected in the CMW context and may indicate a collision system dependence of the LCC or
other mechanisms. The measured normalized slope parameters,
as well as the absolute slope parameters, for each multiplicity or centrality range of \pPb and \PbPb collisions,
are reported in Tables~\ref{tab:pPb}--\ref{tab:PbPb_Cent}.

\begin{table*}[htbp!]
\centering
\topcaption{ \label{tab:pPb} The table summarizes the absolute and normalized slope parameters ($r$) from $v_2$ and $\left< \pt \right>$ in ranges of multiplicity class, $\noff$, in \pPb collisions at \sqrtsNN = 5.02\TeV. The first uncertainty associated with the central values denotes statistical errors, while the second uncertainty represents the systematic uncertainty.}
\cmsTable{\begin{scotch}{lcccc}
 $\noff$	&  $r_{\left< v_2 \right>}$ &  $r^\text{norm}_{\left< v_2 \right>}$	&  $r_{\left< \pt \right>}$ &  $r^\text{norm}_{\left< \pt \right>}$   \\
\hline
$[120,150)$ & 0.022$\pm$0.001$\pm$0.002 & 0.163$\pm$0.01$\pm$0.011 & 0.103$\pm$0.001$\pm$0.007 & 0.06$\pm$0$\pm$0.004 \\
$[150,185)$ & 0.02$\pm$0.001$\pm$0.001 & 0.145$\pm$0.008$\pm$0.009 & 0.105$\pm$0.001$\pm$0.007 & 0.06$\pm$0$\pm$0.004 \\
$[185,220)$ & 0.02$\pm$0.001$\pm$0.001 & 0.143$\pm$0.008$\pm$0.009 & 0.108$\pm$0.001$\pm$0.007 & 0.062$\pm$0.001$\pm$0.004 \\
$[220,260)$ & 0.022$\pm$0.002$\pm$0.001 & 0.153$\pm$0.012$\pm$0.009 & 0.111$\pm$0.002$\pm$0.007 & 0.063$\pm$0.001$\pm$0.004 \\
\end{scotch}}
\end{table*}

\begin{table*}
\centering
\topcaption{ \label{tab:PbPb_Ntrk} The table summarizes the absolute and normalized slope parameters ($r$) from $v_2$ and $\left< \pt \right>$ in ranges of multiplicity class, $\noff$, in \PbPb collisions at \sqrtsNN = 5.02\TeV. The first uncertainty associated with the central values denotes statistical errors, while the second uncertainty represents the systematic uncertainty.}
\cmsTable{
\begin{scotch}{lcccc}
$\noff$	&  $r_{\left< v_2 \right>}$ &  $r^\text{norm}_{\left< v_2 \right>}$	&  $r_{\left< \pt \right>}$ &  $r^\text{norm}_{\left< \pt \right>}$   \\
\hline
$[90,120)$ & 0.02$\pm$0.001$\pm$0.001 & 0.12$\pm$0.007$\pm$0.009 & 0.084$\pm$0.001$\pm$0.006 & 0.056$\pm$0$\pm$0.004 \\
$[120,150)$ & 0.023$\pm$0.001$\pm$0.002 & 0.131$\pm$0.006$\pm$0.009 & 0.084$\pm$0.001$\pm$0.006 & 0.056$\pm$0.001$\pm$0.004 \\
$[150,185)$ & 0.022$\pm$0.001$\pm$0.001 & 0.119$\pm$0.005$\pm$0.008 & 0.087$\pm$0.001$\pm$0.006 & 0.057$\pm$0.001$\pm$0.004 \\
$[185,220)$ & 0.022$\pm$0.001$\pm$0.001 & 0.108$\pm$0.005$\pm$0.007 & 0.087$\pm$0.001$\pm$0.006 & 0.058$\pm$0.001$\pm$0.004 \\
$[220,260)$ & 0.025$\pm$0.001$\pm$0.001 & 0.126$\pm$0.004$\pm$0.008 & 0.091$\pm$0.001$\pm$0.005 & 0.059$\pm$0.001$\pm$0.004 \\
$[260,300)$ & 0.025$\pm$0.001$\pm$0.001 & 0.122$\pm$0.004$\pm$0.007 & 0.093$\pm$0.001$\pm$0.005 & 0.06$\pm$0.001$\pm$0.003 \\
$[300,400)$ & 0.028$\pm$0$\pm$0.001 & 0.133$\pm$0.002$\pm$0.007 & 0.094$\pm$0.001$\pm$0.005 & 0.061$\pm$0$\pm$0.003 \\
$[400,500)$ & 0.03$\pm$0$\pm$0.001 & 0.141$\pm$0.002$\pm$0.007 & 0.099$\pm$0.001$\pm$0.005 & 0.064$\pm$0.001$\pm$0.003 \\
\end{scotch}
}
\end{table*}

\begin{table*}[htbp!]
\centering
\topcaption{ \label{tab:PbPb_Cent} The table summarizes the absolute and normalized slope parameters ($r$) from $v_2$ and $v_{3}$ in ranges of centrality class, in \PbPb collisions at \sqrtsNN = 5.02\TeV. The first uncertainty associated with the central values denotes statistical errors, while the second uncertainty represents the systematic uncertainty.}
\cmsTable{
\begin{scotch}{lcccc}
 Centrality	&  $r_{\left< v_2 \right>}$ &  $r^\text{norm}_{\left< v_2 \right>}$	&  $r_{\left< v_3 \right>}$ &  $r^\text{norm}_{\left< v_3 \right>}$   \\
\hline
30--40\% & 0.032$\pm$0$\pm$0.001 & 0.162$\pm$0.001$\pm$0.006 & 0.01$\pm$0.0006$\pm$0.0004 & 0.149$\pm$0.008$\pm$0.006 \\
40--50\% & 0.032$\pm$0$\pm$0.001 & 0.151$\pm$0.001$\pm$0.006 & 0.0102$\pm$0.0007$\pm$0.0004 & 0.15$\pm$0.01$\pm$0.006 \\
50--60\% & 0.028$\pm$0$\pm$0.001 & 0.135$\pm$0.001$\pm$0.007 & 0.0083$\pm$0.001$\pm$0.0004 & 0.131$\pm$0.016$\pm$0.007 \\
60--70\% & 0.024$\pm$0$\pm$0.002 & 0.126$\pm$0.002$\pm$0.008 & 0.0054$\pm$0.0016$\pm$0.0003 & 0.102$\pm$0.03$\pm$0.006 \\
70--80\% & 0.022$\pm$0.001$\pm$0.002 & 0.136$\pm$0.004$\pm$0.011 & \NA & \NA \\
80--90\% & 0.022$\pm$0.002$\pm$0.002 & 0.171$\pm$0.012$\pm$0.014 & \NA & \NA \\
\end{scotch}
}
\end{table*}

\begin{figure}[htb]
\centering
\includegraphics[width=\cmsFigWidth]{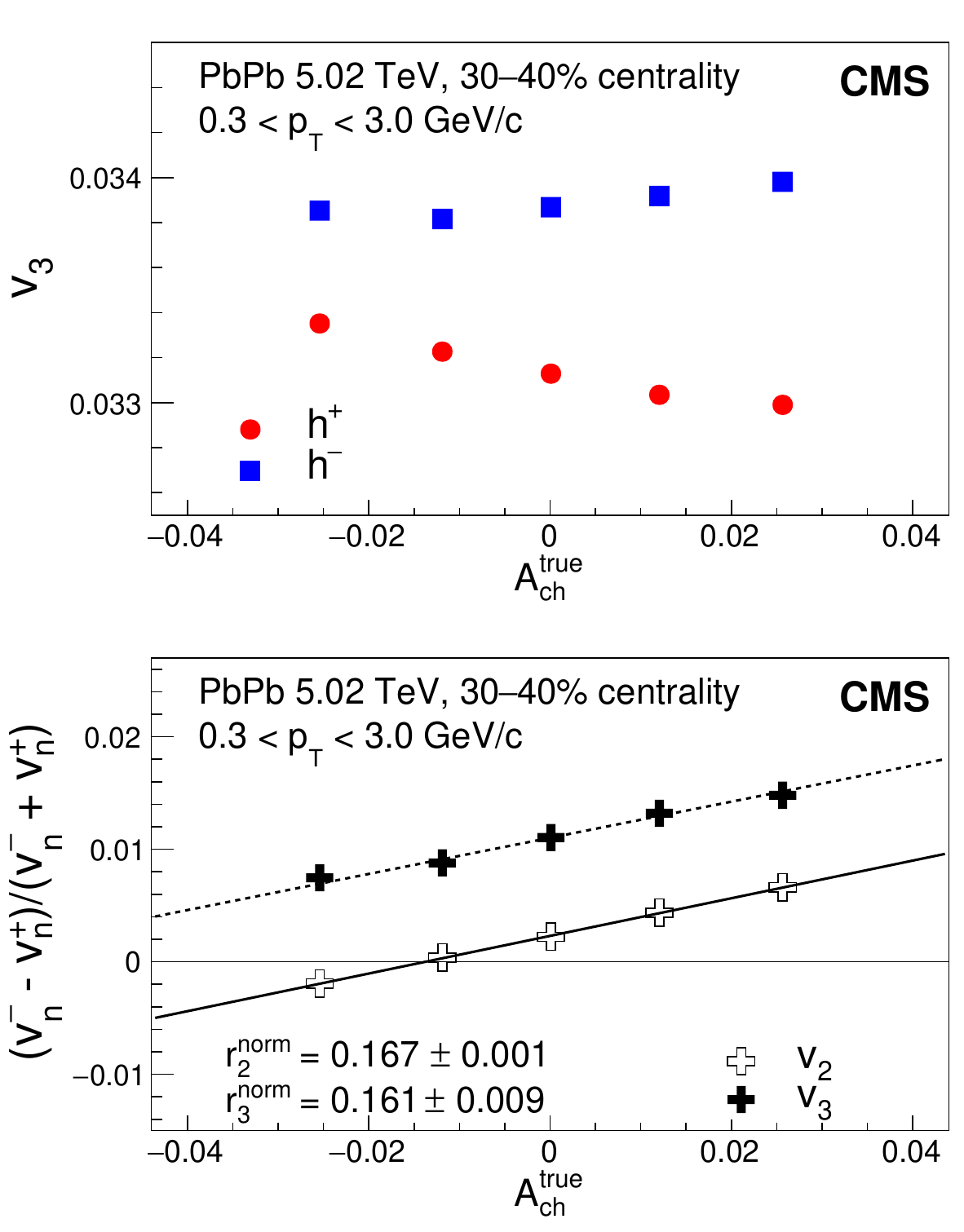}
\caption{
The $v_3$ coefficient for positively and negatively charged particles (top)
and the normalized difference in
$v_{n}$, $(v^{-}_{n} - v^{+}_{n})/(v^{-}_{n} + v^{+}_{n})$ (bottom),
for $n=2$ and 3, as functions of true event charge asymmetry
for the 30--40\% centrality class in  \PbPb collisions at $\sqrtsNN = 5.02$\TeV.
}
\label{fig:figure3}
\end{figure}

The charge asymmetry dependence of the $v_3$ coefficient for positively
and negatively charged particles is also studied in \PbPb
collisions at $\sqrtsNN = 5.02$\TeV, as shown in Fig.~\ref{fig:figure3} (top)
for the 30--40\% centrality class. As found for the $v_2$ values,
the $v^{+}_{3}$ ($v^{-}_{3}$)
values also decrease (increase) as $A^\text{true}_\text{ch}$ increases. No $v_3$ results
for \pPb collisions are reported because of limited statistical precision.
The normalized $v_3$ difference,
$(v^{-}_3-v^{+}_3)/(v^{-}_3+v^{+}_3)$, is derived as
a function of $A^\text{true}_\text{ch}$ in  \PbPb collisions and compared with
that for $v_2$ in Fig.~\ref{fig:figure3} (bottom). The normalized slope
parameter of $v_{3}$, $r^\text{norm}_{3}$, agrees well with $r^\text{norm}_{2}$ within statistical uncertainties.
Charge-dependent higher harmonic $v_\text{n}$ coefficients were measured in \PbPb collisions at 2.76\TeV~\cite{Abelev:2012pa} and their magnitude was found to be smaller than that of the second order coefficient.
We show in this paper that, once normalized, no difference is observed for the $A^\text{true}_\text{ch}$ dependence between the charge-dependent $v_2$ and $v_3$.

\begin{figure}[thb]
\centering
\includegraphics[width=\cmsFigWidth]{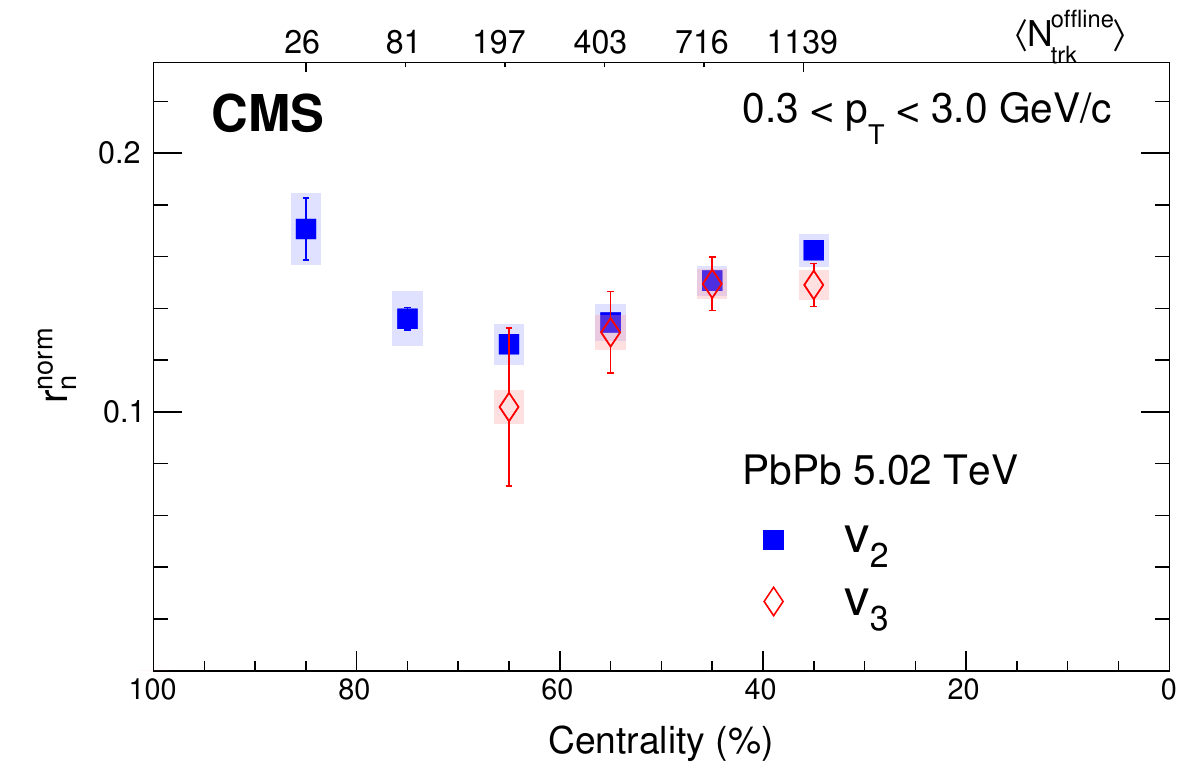}
\caption{
The linear slope parameters, $r^\text{norm}_{2}$ and $r^\text{norm}_{3}$
as functions of the centrality class in  \PbPb collisions.
Average \noff  values for each centrality class are indicated
on the top axis.
Statistical and systematic uncertainties
are indicated by the error bars and shaded regions, respectively.
}
\label{fig:figure4}
\end{figure}

The $r^\text{norm}_{2}$ and $r^\text{norm}_{3}$ values of \PbPb collisions at $\sqrtsNN = 5.02$\TeV\ are shown in Fig.~\ref{fig:figure4}, as functions of centrality in the range 30--90\%. As found for $r^\text{norm}_{2}$, a moderate centrality dependence of $r^\text{norm}_{3}$
is observed. Over the centrality range studied in this analysis, the $r^\text{norm}_{2}$ and
$r^\text{norm}_{3}$ slope parameters are consistent with each other within uncertainties.
The CMW effect is expected with respect to the reaction plane, which is approximated by the second-order event plane in \AonA\ collisions, but highly suppressed with respect to the third-order event plane~\cite{Kharzeev:2015znc}.
The observation of the harmonic order independence, reflected in the similar $r^\text{norm}_{2}$
and $r^\text{norm}_{3}$ values, indicates an underlying physics mechanism unrelated to the CMW effect and, instead,
can be qualitatively explained by the LCC effect~\cite{Bzdak:2013yla}.

Note that the results reported here and elsewhere~\cite{Adam:2015vje, Adamczyk:2015eqo}
used the same population of particles to measure both $v_{n}$ and $A^\text{true}_\text{ch}$.
However, the slope parameters are found to be reduced by about a factor of three, if the $A^\text{true}_\text{ch}$ and $v_{n}$ values are determined by two distinct groups of randomly selected particles. This suggests that the observed correlations are not of a collective nature.

\section{Summary}

In summary, the charge-dependent Fourier coefficients of the azimuthal anisotropy
have been measured in \pPb and \PbPb collisions at $\sqrtsNN = 5.02$\TeV
as functions of the charge asymmetry of the produced hadrons. The normalized differences
in the $v_2$ coefficient between positively and negatively charged particles
in \pPb and \PbPb, and that in the $v_3$ coefficient in \PbPb collisions, are found
to depend linearly on the charge asymmetry. The normalized slope parameters of the
$v_2$ coefficient versus charge asymmetry in \pPb collisions are found to be significant
and similar to those in \PbPb collisions over a wide range of charged particle multiplicities.
The normalized slope parameters of the $v_2$ and $v_3$ coefficients in \PbPb collisions
show similar magnitudes for various centrality classes. A significant charged asymmetry
dependence is also observed for the event-averaged transverse momenta
of positively and negatively charged particles in both \pPb and \PbPb collisions.
None of these observations, made at 5.02\TeV\ and within the CMS phase space window,
are expected from the chiral magnetic wave as the dominant physics mechanism, while
they are qualitatively consistent with predictions based on local charge conservation.
The new measurements presented here indicate that, at LHC energies, the chiral magnetic wave
is not the cause of the charge-dependent azimuthal anisotropies seen in \pPb\ and \PbPb\ collisions.

\begin{acknowledgments}
We congratulate our colleagues in the CERN accelerator departments for the excellent performance of the LHC and thank the technical and administrative staffs at CERN and at other CMS institutes for their contributions to the success of the CMS effort. In addition, we gratefully acknowledge the computing centers and personnel of the Worldwide LHC Computing Grid for delivering so effectively the computing infrastructure essential to our analyses. Finally, we acknowledge the enduring support for the construction and operation of the LHC and the CMS detector provided by the following funding agencies: BMWFW and FWF (Austria); FNRS and FWO (Belgium); CNPq, CAPES, FAPERJ, and FAPESP (Brazil); MES (Bulgaria); CERN; CAS, MoST, and NSFC (China); COLCIENCIAS (Colombia); MSES and CSF (Croatia); RPF (Cyprus); SENESCYT (Ecuador); MoER, ERC IUT, and ERDF (Estonia); Academy of Finland, MEC, and HIP (Finland); CEA and CNRS/IN2P3 (France); BMBF, DFG, and HGF (Germany); GSRT (Greece); OTKA and NIH (Hungary); DAE and DST (India); IPM (Iran); SFI (Ireland); INFN (Italy); MSIP and NRF (Republic of Korea); LAS (Lithuania); MOE and UM (Malaysia); BUAP, CINVESTAV, CONACYT, LNS, SEP, and UASLP-FAI (Mexico); MBIE (New Zealand); PAEC (Pakistan); MSHE and NSC (Poland); FCT (Portugal); JINR (Dubna); MON, RosAtom, RAS, RFBR and RAEP (Russia); MESTD (Serbia); SEIDI, CPAN, PCTI and FEDER (Spain); Swiss Funding Agencies (Switzerland); MST (Taipei); ThEPCenter, IPST, STAR, and NSTDA (Thailand); TUBITAK and TAEK (Turkey); NASU and SFFR (Ukraine); STFC (United Kingdom); DOE and NSF (USA).

\hyphenation{Rachada-pisek} Individuals have received support from the Marie-Curie program and the European Research Council and Horizon 2020 Grant, contract No. 675440 (European Union); the Leventis Foundation; the A. P. Sloan Foundation; the Alexander von Humboldt Foundation; the Belgian Federal Science Policy Office; the Fonds pour la Formation \`a la Recherche dans l'Industrie et dans l'Agriculture (FRIA-Belgium); the Agentschap voor Innovatie door Wetenschap en Technologie (IWT-Belgium); the Ministry of Education, Youth and Sports (MEYS) of the Czech Republic; the Council of Science and Industrial Research, India; the HOMING PLUS program of the Foundation for Polish Science, cofinanced from European Union, Regional Development Fund, the Mobility Plus program of the Ministry of Science and Higher Education, the National Science Center (Poland), contracts Harmonia 2014/14/M/ST2/00428, Opus 2014/13/B/ST2/02543, 2014/15/B/ST2/03998, and 2015/19/B/ST2/02861, Sonata-bis 2012/07/E/ST2/01406; the National Priorities Research Program by Qatar National Research Fund; the Programa Clar\'in-COFUND del Principado de Asturias; the Thalis and Aristeia programs cofinanced by EU-ESF and the Greek NSRF; the Rachadapisek Sompot Fund for Postdoctoral Fellowship, Chulalongkorn University and the Chulalongkorn Academic into Its 2nd Century Project Advancement Project (Thailand); the Welch Foundation, contract C-1845; and the Weston Havens Foundation (USA). \end{acknowledgments}

\bibliography{auto_generated}
\cleardoublepage \appendix\section{The CMS Collaboration \label{app:collab}}\begin{sloppypar}\hyphenpenalty=5000\widowpenalty=500\clubpenalty=5000\vskip\cmsinstskip
\textbf{Yerevan~Physics~Institute,~Yerevan,~Armenia}\\*[0pt]
A.M.~Sirunyan, A.~Tumasyan
\vskip\cmsinstskip
\textbf{Institut~f\"{u}r~Hochenergiephysik,~Wien,~Austria}\\*[0pt]
W.~Adam, F.~Ambrogi, E.~Asilar, T.~Bergauer, J.~Brandstetter, E.~Brondolin, M.~Dragicevic, J.~Er\"{o}, M.~Flechl, M.~Friedl, R.~Fr\"{u}hwirth\cmsAuthorMark{1}, V.M.~Ghete, J.~Grossmann, J.~Hrubec, M.~Jeitler\cmsAuthorMark{1}, A.~K\"{o}nig, N.~Krammer, I.~Kr\"{a}tschmer, D.~Liko, T.~Madlener, I.~Mikulec, E.~Pree, D.~Rabady, N.~Rad, H.~Rohringer, J.~Schieck\cmsAuthorMark{1}, R.~Sch\"{o}fbeck, M.~Spanring, D.~Spitzbart, W.~Waltenberger, J.~Wittmann, C.-E.~Wulz\cmsAuthorMark{1}, M.~Zarucki
% ...
\vskip\cmsinstskip
\dag:~Deceased\\
d:~Also at~Vienna~University~of~Technology,~Vienna,~Austria\\
d:~Also at~State~Key~Laboratory~of~Nuclear~Physics~and~Technology;~Peking~University,~Beijing,~China\\
d:~Also at~Universidade~Estadual~de~Campinas,~Campinas,~Brazil\\
d:~Also at~Universidade~Federal~de~Pelotas,~Pelotas,~Brazil\\
d:~Also at~Universit\'{e}~Libre~de~Bruxelles,~Bruxelles,~Belgium\\
d:~Also at~Institute~for~Theoretical~and~Experimental~Physics,~Moscow,~Russia\\
d:~Also at~Joint~Institute~for~Nuclear~Research,~Dubna,~Russia\\
d:~Also at~Suez~University,~Suez,~Egypt\\
d:~Now at~British~University~in~Egypt,~Cairo,~Egypt\\
d:~Also at~Fayoum~University,~El-Fayoum,~Egypt\\
d:~Now at~Helwan~University,~Cairo,~Egypt\\
d:~Also at~Universit\'{e}~de~Haute~Alsace,~Mulhouse,~France\\
d:~Also at~Skobeltsyn~Institute~of~Nuclear~Physics;~Lomonosov~Moscow~State~University,~Moscow,~Russia\\
d:~Also at~Tbilisi~State~University,~Tbilisi,~Georgia\\
d:~Also at~Ilia~State~University,~Tbilisi,~Georgia\\
d:~Also at~CERN;~European~Organization~for~Nuclear~Research,~Geneva,~Switzerland\\
d:~Also at~RWTH~Aachen~University;~III.~Physikalisches~Institut~A,~Aachen,~Germany\\
d:~Also at~University~of~Hamburg,~Hamburg,~Germany\\
d:~Also at~Brandenburg~University~of~Technology,~Cottbus,~Germany\\
d:~Also at~MTA-ELTE~Lend\"{u}let~CMS~Particle~and~Nuclear~Physics~Group;~E\"{o}tv\"{o}s~Lor\'{a}nd~University,~Budapest,~Hungary\\
d:~Also at~Institute~of~Nuclear~Research~ATOMKI,~Debrecen,~Hungary\\
d:~Also at~Institute~of~Physics;~University~of~Debrecen,~Debrecen,~Hungary\\
d:~Also at~Indian~Institute~of~Technology~Bhubaneswar,~Bhubaneswar,~India\\
d:~Also at~Institute~of~Physics,~Bhubaneswar,~India\\
d:~Also at~University~of~Visva-Bharati,~Santiniketan,~India\\
d:~Also at~University~of~Ruhuna,~Matara,~Sri~Lanka\\
d:~Also at~Isfahan~University~of~Technology,~Isfahan,~Iran\\
d:~Also at~Yazd~University,~Yazd,~Iran\\
d:~Also at~Plasma~Physics~Research~Center;~Science~and~Research~Branch;~Islamic~Azad~University,~Tehran,~Iran\\
d:~Also at~Universit\`{a}~degli~Studi~di~Siena,~Siena,~Italy\\
d:~Also at~INFN~Sezione~di~Milano-Bicocca;~Universit\`{a}~di~Milano-Bicocca,~Milano,~Italy\\
d:~Also at~Laboratori~Nazionali~di~Legnaro~dell'INFN,~Legnaro,~Italy\\
d:~Also at~Purdue~University,~West~Lafayette,~USA\\
d:~Also at~International~Islamic~University~of~Malaysia,~Kuala~Lumpur,~Malaysia\\
d:~Also at~Malaysian~Nuclear~Agency;~MOSTI,~Kajang,~Malaysia\\
d:~Also at~Consejo~Nacional~de~Ciencia~y~Tecnolog\'{i}a,~Mexico~city,~Mexico\\
d:~Also at~Warsaw~University~of~Technology;~Institute~of~Electronic~Systems,~Warsaw,~Poland\\
d:~Also at~Institute~for~Nuclear~Research,~Moscow,~Russia\\
d:~Now at~National~Research~Nuclear~University~'Moscow~Engineering~Physics~Institute'~(MEPhI),~Moscow,~Russia\\
d:~Also at~St.~Petersburg~State~Polytechnical~University,~St.~Petersburg,~Russia\\
d:~Also at~University~of~Florida,~Gainesville,~USA\\
d:~Also at~P.N.~Lebedev~Physical~Institute,~Moscow,~Russia\\
d:~Also at~Budker~Institute~of~Nuclear~Physics,~Novosibirsk,~Russia\\
d:~Also at~Faculty~of~Physics;~University~of~Belgrade,~Belgrade,~Serbia\\
d:~Also at~University~of~Belgrade;~Faculty~of~Physics~and~Vinca~Institute~of~Nuclear~Sciences,~Belgrade,~Serbia\\
d:~Also at~Scuola~Normale~e~Sezione~dell'INFN,~Pisa,~Italy\\
d:~Also at~National~and~Kapodistrian~University~of~Athens,~Athens,~Greece\\
d:~Also at~Riga~Technical~University,~Riga,~Latvia\\
d:~Also at~Universit\"{a}t~Z\"{u}rich,~Zurich,~Switzerland\\
d:~Also at~Stefan~Meyer~Institute~for~Subatomic~Physics~(SMI),~Vienna,~Austria\\
d:~Also at~Adiyaman~University,~Adiyaman,~Turkey\\
d:~Also at~Istanbul~Aydin~University,~Istanbul,~Turkey\\
d:~Also at~Mersin~University,~Mersin,~Turkey\\
d:~Also at~Cag~University,~Mersin,~Turkey\\
d:~Also at~Piri~Reis~University,~Istanbul,~Turkey\\
d:~Also at~Izmir~Institute~of~Technology,~Izmir,~Turkey\\
d:~Also at~Necmettin~Erbakan~University,~Konya,~Turkey\\
d:~Also at~Marmara~University,~Istanbul,~Turkey\\
d:~Also at~Kafkas~University,~Kars,~Turkey\\
d:~Also at~Istanbul~Bilgi~University,~Istanbul,~Turkey\\
d:~Also at~Rutherford~Appleton~Laboratory,~Didcot,~United~Kingdom\\
d:~Also at~School~of~Physics~and~Astronomy;~University~of~Southampton,~Southampton,~United~Kingdom\\
d:~Also at~Instituto~de~Astrof\'{i}sica~de~Canarias,~La~Laguna,~Spain\\
d:~Also at~Utah~Valley~University,~Orem,~USA\\
d:~Also at~Beykent~University,~Istanbul,~Turkey\\
d:~Also at~Bingol~University,~Bingol,~Turkey\\
d:~Also at~Erzincan~University,~Erzincan,~Turkey\\
d:~Also at~Sinop~University,~Sinop,~Turkey\\
d:~Also at~Mimar~Sinan~University;~Istanbul,~Istanbul,~Turkey\\
d:~Also at~Texas~A\&M~University~at~Qatar,~Doha,~Qatar\\
d:~Also at~Kyungpook~National~University,~Daegu,~Korea\\
\end{sloppypar}
\end{document}